\documentclass{siam-wns-article}
\def\tr{\mathop{\mathrm{Tr}}}
\newtheorem{theorem}{Theorem}

\newtheorem{conjecture}[theorem]{Conjecture}
\title{Spectral bounds for percolation on directed and undirected graphs}

\author{Kathleen E. Hamilton and Leonid P. Pryadko}

\usehyperref 

\begin{document}
\maketitle

\section*{Summary}
We suggest several algebraic bounds for percolation on directed and
undirected graphs: proliferation of strongly-connected clusters,
proliferation of in- and out-clusters, and the transition associated
with the number of giant components.  

\section*{Introduction}
Percolation on random graphs has been successfully used in network
theory as a way to understand connectivity of real-life
networks\cite{bollobas1998random,%
  Callaway-Newman-Strogatz-Watts-2000,Hofstad-2010}.  On a tree graph,
mean field theory is useful to determine the location of a percolation
transition \cite{stauffer1991introduction}. However the existence of
cycles reduce the applicability of this method on general graphs. We
seek rigorous mean field bounds on the percolation transition which
would be applicable to any graph.

Spectra of associated matrices have been linked to
percolation and epidemic thresholds for a variety of
graph and network models
\cite{PhysRevE.76.056119,Restrepo-Ott-Hunt-2008,chakrabarti2008epidemic,%
  Bollobas-Borgs-Chayes-Riordan-2010,prakash2012threshold}. Recently,
we constructed a lower bound on the percolation transition for an
infinite quasi-transitive graph $\mathcal{G}_0$, a graph-theoretical
analog of a translationally-invariant
system\cite{Hamilton-Pryadko-PRL-2014}.  The bound,
\begin{equation}
  \label{eq:orig-bound}
  p_c\ge 1/\rho(H),
\end{equation}
is defined by the inverse spectral radius of the Hashimoto
matrix\cite{Hashimoto-matrix-1989} $H$ which is used to enumerate
non-backtracking walks on $\mathcal{G}_0$.  The formal proof involved
a sequential application of \emph{cycle unwrapping} maps which results
in a tree graph  locally equivalent to the original graph. In
the case of a degree-regular graph, Eq.~(\ref{eq:orig-bound}) gives
the well-known bound in terms of the maximum degree.  In the case of
random graphs with few short cycles, the same expression
(\ref{eq:orig-bound}) gives a numerically exact result for the
percolation transition\cite{Karrer-Newman-Zdeborova-PRL-2014}.

In this work we analyze the applicability of the expression
(\ref{eq:orig-bound}) to more general graphs and in particular, finite
graphs.  We note that the Hashimoto matrix $H$ can be viewed as the
adjacency matrix of the oriented line (di)graph (OLG) associated with
the original graph\cite{Kotani-Sunada-2000}.  Thus,
Eq.~(\ref{eq:orig-bound}) requires a mapping of the original
\emph{undirected} percolation problem to a \emph{directed} one.
Several transitions are associated with percolation on a digraph
${\cal D}$, in particular, formation of a giant strongly connected
component, and formation of a giant in- or out-
component\cite{Restrepo-Ott-Hunt-2008}.  We show that on a general
digraph, Eq.~(\ref{eq:orig-bound}) is most directly associated with
another transition, the formation of a strongly connected component
with a large number of distinct self-avoiding cycles (SACs), and argue
that this property is related to uniqueness of the percolating
cluster.

\section{Main results}
Consider an order-$n$
strongly-con\-nected digraph ${\cal D}$ with vertex and edge sets
${\cal V}({\cal D})$, ${\cal E}({\cal D})$. Associated with ${\cal D}$
is an adjacency matrix $A\equiv A({\cal D})$, and Hashimoto
matrix\cite{Hashimoto-matrix-1989} $H\equiv H({\cal D})$, with
elements%
\begin{equation}
  \label{eq:Hashimoto}
  H_{u,v}=\delta_{jj'}(1-\delta_{li}),\quad u\equiv i\to j,\;v\equiv
  j'\to l, 
\end{equation}
where $u$, $v$ are directed edges in ${\cal E}({\cal D})$.  The
digraph ${\cal D}$ may have some symmetric edges (length-two cycles),
whereas the OLG has none.  The spectral
radius of $H$ satisfies $\rho(H)\le \rho(A)$; the equality
is reached iff ${\cal D}$ has no symmetric edges.  Also, for any
induced $q$-norm, $\rho(H)\le \|H\|_q$.

In \emph{site percolation} on ${\cal D}$, each vertex is open with
probability $p$ and closed with probability $1-p$; we consider a
subgraph ${\cal D}'$ of ${\cal D}$ induced by the open vertices.  We
are interested in the likelihood of forming in-, out-, or strongly
connected components of a given size $m$ on ${\cal D}'$ [e.g., in the
case of the out-component, there exists a vertex $i_0\in{\cal V}({\cal
  D}')$ such that $m-1$ or more vertices can be reached by directed
walks on ${\cal D}'$ starting with $i_0$].  In a strongly connected
component, each site can be reached from any other by directed walks;
it is automatically both an in- and out-cluster.  We prove the
following:
\begin{theorem}
  \label{th:H-norm-one-bound}
  For any finite $m>0$, the probability that a site $v$ is the
  root of an out-component of size $m$ or greater is bounded by $m
  P_{m}^\mathrm{(out)}(v)\le(1-p\|H\|_1)^{-1}$, $p\|H\|_1<1$.
\end{theorem}
This implies that out-cluster percolation threshold in large
(di)graphs satisfies $p_c^\mathrm{(out)}>\|H\|_1^{-1}$.  A similar
statement can be written for in-cluster percolation,
$p_c^\mathrm{(in)}\ge \|H^T\|_1^{-1}$.  This is a generalization of the
maximum degree bound to arbitrary digraphs.  The proof is based
on constructing an upper bound for the average number of sites
reachable from $v$ via non-backtracking walks.  This is found by
upper-bounding the sum $\|\sum_{m\ge0} p^m H^m x_v\|_1$, where $x_v$
has only one non-zero component.  
We constructed a family of strongly-connected digraphs which saturate
this bound, while at the same time $\rho(H)<\|H\|_1$.  Additionally,
on these digraphs, the strongly-connected cluster percolation
threshold is strictly higher than $p_c^{(\rm out)}$ or $p_c^{(\rm
  in)}$; they give a counterexample for the central conjecture of
Ref.~\cite{Restrepo-Ott-Hunt-2008}.

We notice that OLG of a strongly connected digraph $\mathcal{D}$ is
also strongly connected, provided that $\mathcal{D}$ remains strongly
connected when any edge in any pair $u=i\to j$, $\bar u=j\to i$ of
mutually opposite edges (if any) is removed.  For such a digraph, an
improved bound for out-cluster probability reads $m
P_m^\mathrm{(out)}(v)\le \gamma_L [1-p\rho(H)]^{-1}$ (cf.\ Theorem
\ref{th:H-norm-one-bound}), where $\gamma_L$ is the principal ratio,
$\gamma_L\equiv\max_{ij}(\xi_i/\xi_j)$, for the left Perron-Frobenius
vector of $H$, $\xi\rho(H)=\xi H$.  This inequality allows to extend
the bound (\ref{eq:orig-bound}) from
Ref.~\cite{Hamilton-Pryadko-PRL-2014} to in- and out-cluster
percolation on any strongly-connected quasi-transitive infinite
digraph.

The trace $s^{-1}\tr H^s$ counts non-backtracking directed cycles of
length $s$ on $\mathcal{D}$.  We constructed a corresponding bound for
the total number of SACs on ${\cal D}'$, $n_E\equiv
  |\mathcal{E}(\mathcal{D})|$:%
\begin{equation}
  N\le \sum_{s>0} s^{-1}p^s
  \tr H^s\le n_E\left|\ln[1-p\rho(H)]\right|.
\label{eq:SAC-bound}
\end{equation}
Notice that a strongly-connected cluster with $m$ cycles supports $m$
distinct SACs iff it has an articulation point separating any two
cycles; such a cluster can be separated into two by removing a single
site.  A two-site-connected cluster (which can be cut into two pieces
by removing two sites, but not one) will support at least $m(m-1)/2$
distinct SACs.  A \emph{stable} giant cluster which can not be easily
separated into two large pieces will necessarily have large
connectivity and support up to the maximum $\mathcal{O}(2^m)$ SACs.
For a cluster occupying a finite fraction of directed edges in
$\mathcal{D}'$, such a possibility is excluded by
Eq.~(\ref{eq:SAC-bound}).  Therefore, we expect any large
clusters formed at $p\rho(H)<1$ to be unstable; i.e., fracture easily
into two or more large cluster.  Thus, we give
\begin{conjecture} \label{th:adjacency} 
   For $p\rho(H)<1$, if there is a percolating cluster, it is not
   unique with probability approaching one at large $n$.
\end{conjecture}

\section{Conclusions} 
We used algebraic techniques based on non-backtracking (Hashimoto)
matrix to establish several new bounds for percolation on general
directed (and undirected) graphs.  Our results are formulated as
bounds for the probability that a given site connects to a cluster of
size $m$ or greater, and are
applicable for finite and infinite (di)graphs.  We also discuss the
stability of giant strongly-connected clusters, which is related to
the uniqueness transition.

\section{Acknowledgements}
The authors wish to thank Mark E. J. Newman for insightful
discussions.  This work was supported in part by the U.S. Army
Research Office under Grant No.  W911NF-14-1-0272 and by the NSF under
Grant No. PHY-1416578.  

\begin{thebibliography}{10}

\bibitem{bollobas1998random}
B.~Bollob{\'a}s.
\newblock {\em Random graphs}.
\newblock Springer, 1998.

\bibitem{Bollobas-Borgs-Chayes-Riordan-2010}
B.~Bollob{\'{a}}s, C.~Borgs, J.~Chayes, and O.~Riordan.
\newblock Percolation on dense graph sequences.
\newblock {\em The Annals of Probability}, 38(1):150--183, 2010.

\bibitem{Callaway-Newman-Strogatz-Watts-2000}
D.~S. Callaway, M.~E.~J. Newman, S.~H. Strogatz, and D.~J. Watts.
\newblock Network robustness and fragility: Percolation on random graphs.
\newblock {\em Phys. Rev. Lett.}, 85:5468--5471, Dec 2000.

\bibitem{chakrabarti2008epidemic}
D.~Chakrabarti, Y.~Wang, C.~Wang, J.~Leskovec, and C.~Faloutsos.
\newblock Epidemic thresholds in real networks.
\newblock {\em ACM Transactions on Information and System Security (TISSEC)},
  10(4):1, 2008.

\bibitem{Hamilton-Pryadko-PRL-2014}
K.~E. Hamilton and L.~P. Pryadko.
\newblock Tight lower bound for percolation threshold on an infinite graph.
\newblock {\em Phys. Rev. Lett.}, 113:208701, Nov 2014.

\bibitem{Hashimoto-matrix-1989}
K.~Hashimoto.
\newblock Zeta functions of finite graphs and representations of $p$-adic
  groups.
\newblock In K.~Hashimoto and Y.~Namikawa, editors, {\em Automorphic Forms and
  Geometry of Arithmetic Varieties}, volume~15 of {\em Advanced Studies in Pure
  Mathematics}, pages 211--280. Kinokuniya, Tokyo, 1989.

\bibitem{Karrer-Newman-Zdeborova-PRL-2014}
B.~Karrer, M.~E.~J. Newman, and L.~Zdeborov\'a.
\newblock Percolation on sparse networks.
\newblock {\em Phys. Rev. Lett.}, 113:208702, Nov 2014.

\bibitem{Kotani-Sunada-2000}
M.~Kotani and T.~Sunada.
\newblock Zeta functions of finite graphs.
\newblock {\em J. Math. Sci. Univ. Tokyo}, 7(1):7--25, 2000.

\bibitem{prakash2012threshold}
B.~A. Prakash, D.~Chakrabarti, N.~C. Valler, M.~Faloutsos, and C.~Faloutsos.
\newblock Threshold conditions for arbitrary cascade models on arbitrary
  networks.
\newblock {\em Knowledge and information systems}, 33(3):549--575, 2012.

\bibitem{PhysRevE.76.056119}
J.~G. Restrepo, E.~Ott, and B.~R. Hunt.
\newblock Approximating the largest eigenvalue of network adjacency matrices.
\newblock {\em Phys. Rev. E}, 76:056119, Nov 2007.

\bibitem{Restrepo-Ott-Hunt-2008}
J.~G. Restrepo, E.~Ott, and B.~R. Hunt.
\newblock Weighted percolation on directed networks.
\newblock {\em Phys. Rev. Lett.}, 100:058701, Feb 2008.

\bibitem{stauffer1991introduction}
D.~Stauffer and A.~Aharony.
\newblock {\em Introduction to percolation theory}.
\newblock Taylor and Francis, 1991.

\bibitem{Hofstad-2010}
R.~van~der Hofstad.
\newblock Percolation and random graphs.
\newblock In I.~Molchanov and W.~Kendall, editors, {\em New Perspectives on
  Stochastic Geometry}, chapter~6, pages 173--247. Oxford University Press,
  2010.
\newblock ISBN 978-0-19-923257-4.

\end{thebibliography}

\end{document}